\documentclass{article}%
\usepackage{wrapft}
\usepackage{amsmath}
\usepackage{amsfonts}
\usepackage{amssymb}
\usepackage{graphicx}%
\usepackage{here}
\usepackage{color}
\usepackage[super,compress]{cite}
\usepackage{geometry}
\begin{document}
%


%
%

\title{Dark matter in logarithmic $F(R)$ gravity}

\author{
Tomohiro Inagaki$^{1,2,3,\dagger}$ ,Yamato Matsuo$^{4,\ddagger}$ and Hiroki Sakamoto$^{4,\ast}$\\
\date{
$^1$Information Media Center, Hiroshima University, Higashi-Hiroshima, 739-8521, Japan, \\
$^2$Core of Research for the Energetic Universe, Hiroshima University, Higashi-Hiroshima, 739-8526, Japan\\
$^3$Lab. Theor. Cosmology, Tomsk State University of Control Systems and Radioelectronics (TUSUR), 634050 Tomsk, Russia,\\
$^4$Department of Physics, Hiroshima University, Higashi-Hiroshima, 739-8526, Japan \\
$^\dagger$inagaki@hiroshima-u.ac.jp\\
$^\ddagger$ya-matsuo@hiroshima-u.ac.jp\\
$^\ast$h-sakamoto@hiroshima-u.ac.jp\\
}
}


%

%
%

\maketitle


\begin{abstract}
The logarithmic $R^2$-corrected $F(R)$ gravity is investigated as a prototype model of modified gravity theories with quantum corrections. By using the auxiliary field method, the model is described by the general relativity with a scalaron field. The scalaron field can be identified as an inflaton at the primordial inflation era. It is also one of the dark matter candidates in the dark energy era. It is found that a wide range of the parameters is consistent with the current observations of CMB fluctuations, dark energy and dark matter.
\end{abstract}

%

\section{Introduction}

$F(R)$ gravity is one of the interesting extensions of the general relativity. As is known, a class of $F(R)$ gravity induces the primordial and current accelerated expansion of the universe, for a review, see Refs.~\citen{Nojiri:2006ri,Nojiri:2010wj,Clifton:2011jh,Nojiri:2017ncd} and references therein. The model reproduces the general relativity at the weak curvature limit and has the potential to solve the origin of the primordial inflation and dark energy. The additional physical degree of freedom in the $F(R)$ gravity can be described by a scalar field. 

The fluctuation of the scalar field around the minimum of the effective potential can be regarded as a massive scalaron field. Because of the chameleon mechanism the scalaron field is light enough in a galaxy scale while it is too heavy to observe in high energy experiments on the earth. It is pointed out that the scalaron field can play a role of the dark matter in $R^2$ modified gravity \cite{Cembranos:2008gj,Nojiri:2008nt,Cembranos:2010qd}. In Refs.~\citen{Katsuragawa:2016yir,Katsuragawa:2017wge} it is found that fine-tuning of the model parameter or introducing an additional term is necessary to stabilize the scalar field against the two-photon decay in the Starobinsky like model \cite{Starobinsky:2007hu}.

It is considered that leading order corrections in a higher derivative quantum gravity introduce logarithmic terms with respect to the curvature $R$. Hence, the logarithmic $F(R)$ gravity is regarded as a prototype model of the gravity theory with quantum corrections and able to describe the primordial and current accelerated expansion of the universe \cite{Odintsov:2017hbk}. In this paper we consider the scalaron field in  the logarithmic $F(R)$ gravity as a dark matter candidate and investigate the allowed region for parameters of the logarithmic $F(R)$ gravity at the inflation era and dark energy era.

The paper is organized as follows. In Sec.~2 we introduce the logarithmic $R^2$-corrected $F(R)$ gravity and derive the equivalent scalar-tensor theory. In Sec.~3 we consider the scalaron field at the dark energy (DE) era and evaluate the allowed parameter range as a dark matter candidate. The section 4 is devoted for the concluding remarks.
\section{Logarithmic $F(R)$ gravity}

\subsection{Set up}
Throughout this paper we employ the logarithmic $R^2$-corrected $F(R)$ gravity defined by the action
\begin{align}
  S=\frac{1}{2\kappa^2}\int d^4x\sqrt{-g}F(R)+S_{Matter},\label{F(R)action}
\end{align}
with
\begin{align}
  F (R)  =  R - \Lambda_{DE} \left[ 1 - \alpha \frac{R}{R_C} \ln
  \left( \frac{R}{R_C} \right) \right] + \kappa^2 \gamma_0 \left[ 1 + \gamma_1
  \ln \left( \frac{R}{R_0} \right) \right] R^2, \label{logF(R)}
\end{align}
where $\kappa$ is inverse of the reduced Planck mass and $S_{Matter}$ represents the action of Standard Model (SM) particles, $\Lambda_{DE}$ describes the current  cosmological constant. The model has five free constant parameters, $\alpha$, $R_C$, $\gamma_0$, $\gamma_1$ and $R_0$. Here we assume the parameter $R_C$ at the DE scale and $R_0$ at the primordial inflation scale.

This model is based on the Starobinsky type $R^2$ modified gravity,
\begin{align}
  F(R) = R +\Lambda_{DE}+\kappa^2\gamma R^2.
\end{align}
If we consider leading order quantum corrections in a multiplicatively renormalizable higher-derivative quantum gravity, the cosmological constant, $\Lambda_{DE}$, and the coefficient, $\gamma$, in front of $R^2$ obtain logarithmic corrections as Eq.(\ref{logF(R)}) \cite{Odintsov:2017hbk}.

For practical calculations it is more convenient to introduce an auxiliary field $A$ and rewrite the action (\ref{F(R)action}) as
\begin{align}
  S=\frac{1}{2\kappa^2}\int d^4x\sqrt{-g}\left[
  F'(A)R-\{
  F'(A)A-F(A)
  \}
  \right]
  +S_{Matter}.\label{F(R)auxiliary}
\end{align}
From this action the equation of motion for $A$ is given by
\begin{align}
  F''(A)(R-A)=0,
\end{align}
where $F'(A)$ and $F''(A)$ represents the first and second derivatives of $F(A)$.
The original action (\ref{F(R)action}) is reproduced by substituting the solution of the equation of motion, $A=R$, to the action (\ref{F(R)auxiliary}).


\subsection{Scalaron description}

The action (\ref{F(R)auxiliary}) is a kind of scalar-tensor theory in the Jordan frame. The theory can be expressed by an equivalent action in the Einstein frame. Performing the Weyl transformation, $g_{\mu\nu}\rightarrow\tilde g_{\mu\nu}=F'(A)g_{\mu\nu}$ and replacing the auxiliary field, $A$, with the scalaron field, $\varphi$,
\begin{align}
  F'(A)\equiv e^{2\kappa\varphi/\sqrt6},
  \label{Rtophi}
\end{align}
we obtain the action in the Einstein frame,
\begin{align}
  S=\frac{1}{2\kappa^2}\int d^4x\sqrt{-\tilde g}\tilde R+\int d^4x\sqrt{-\tilde g}\left[
  -\frac12\tilde g^{\mu\nu}(\partial_\mu\varphi)(\partial_\nu\varphi)-V(\varphi)
  \right]
  +\tilde S_{Matter},
\end{align}
where $\tilde R$ and $\tilde S_{Matter}$ describe the Ricci scalar and the SM action with respect to the transformed metric, $\tilde g_{\mu\nu}$. The potential $V(\varphi)$ is given by
\begin{align}
  V(\varphi) = \frac{1}{2\kappa^2}\frac{F'(A)A-F(A)}{F'^2(A)} .
  \label{pot}
\end{align}
From Eq.~(\ref{Rtophi}) and the solution of the equation of motion, $A=R$, the scalaron field $\varphi$ is given as a function of the Ricci scalar. 

The scalaron field interacts with the SM particles in the action, $\tilde S_{Matter}$. Thus the energy-momentum tensor of SM, $T^{\mu\nu}$, contributes the equation of motion for the scalaron field. The contribution is included in the effective potential,
\begin{align}
  V_{eff}(\varphi)\equiv V(\varphi)-\frac{1}{4}e^{-4\kappa\varphi/\sqrt6}{T^\mu}_\mu.
  \label{epot}
\end{align}
If the effective potential is bounded below, the scalaron field has a stable ground state. Then, the scalaron mass is defined by the second derivative of the effective potential at the minimum,
\begin{align}
\left.\frac{d^2 V_{eff}(\varphi)}{d \varphi^2}\right|_{\varphi=\varphi_{min}}=m_\varphi^2.
\label{mass}
\end{align}
Since any consequences of the scalaron field have not been observed in the laboratory, the interactions between the scalaron and SM particles should be weak enough at the solar system scale. From Eq.(\ref{epot}) it is found that the scalaron mass depends on the trace of the energy-momentum tensor. Since this dependence generates large scalaron mass at the solar system scale, the interaction between the scalaron and SM particles is screened out. It is known as the chameleon mechanism. \cite{Katsuragawa:2016yir,Katsuragawa:2017wge,Khoury:2003rn,Brax:2008hh}.


%

\subsection{Constraints at inflation era}

The third term in Eq.~(\ref{logF(R)}) induces exponential expansion of the universe at the inflation era, $R\sim R_0 $. The allowed parameters range is systematically studied in Ref.~\citen{Odintsov:2017hbk}. To reproduce the inflationary parameters obtained in the Starobinsky $R^2$ modified gravity model, the second term in Eq.~(\ref{logF(R)}) should be much smaller than the other terms. Comparing the first and second terms at the scale $R\sim R_0 \gg R_C\sim \Lambda_{DE}>0$, we can ignore the second term for
\begin{align}
  R_0 \gg \alpha \Lambda_{DE} \frac{R_0}{R_C}\ln (R_0 / R_C) .
  \label{cond:inf:1}
\end{align}
Next we also assume $\ln (R/R_0)\sim O(1)$ and compare the second term and the logarithmic correction for the third term in Eq.~(\ref{logF(R)}). The contribution from the second term can be ignored for
\begin{align}
R_0^2   \gg \frac{1}{ \kappa^2\gamma_0 \gamma_1} \Lambda_{DE} \left[1-\alpha \frac{R_0}{R_C}\ln (R_0 / R_C)\right] .
\label{ineq:inf}
\end{align}
The inequality (\ref{ineq:inf}) is obviously satisfied for
\begin{align}
 & R_0^2   \gg \frac{1}{\kappa^2 \gamma_0 \gamma_1}\Lambda_{DE} ,   \label{cond:inf:2}\\
 & R_0^2   \gg \frac{1}{ \kappa^2\gamma_0 \gamma_1}\alpha \Lambda_{DE} \frac{R_0}{R_C}\ln (R_0 / R_C) .
   \label{cond:inf:3}
\end{align}
Under the conditions (\ref{cond:inf:1}), (\ref{cond:inf:2}) and (\ref{cond:inf:3}) the second term in Eq.~(\ref{logF(R)}) is ignored in the inflation era. Then the $F(R)$ reads
\begin{align}
  F (R)  \sim  R + \kappa^2 \gamma_0 \left[ 1 + \gamma_1
  \ln \left( \frac{R}{R_0} \right) \right] R^2. \label{logF(R):inf}
\end{align}

From Eqs.~(\ref{Rtophi}),~(\ref{logF(R):inf}) and the solution of the equation of motion, $A=R$, the Ricci scalar $R$ is given as a function of the scalaron field, $\varphi$, 
\begin{align}
  R(\varphi) = &
  \frac{e^{2\kappa\varphi/\sqrt6}-1}{2\kappa^2\gamma_0\gamma_1W(G(\varphi))},\label{Rtophi-log-inf}
\end{align}
with
\begin{align}
  G(\varphi)
  = &
    \frac{
    e^{\frac{1}{2}+\frac{1}{\gamma_1}}
    \left(
      e^{2\kappa\varphi/\sqrt6}-1
    \right)}
    {2\kappa^2\gamma_0\gamma_1R_0}.
\end{align}
In Eq.~(\ref{Rtophi-log-inf}) we use the Lambelt $W$-function which is defined by the inverse of the following function,
\begin{align}
  z=W(z)e^{W(z)}.
\end{align}
Substituting Eq.~(\ref{logF(R):inf}) into Eq.~(\ref{pot}), we obtain the scalaron potential in the inflation era,
\begin{align}
  V(\varphi) = &
 \frac{1}{2\kappa^2}\frac{\kappa^2\gamma_0}{F(R(\varphi))'^2}
 \left[
   1+\gamma_1+\gamma_1\ln
   \left(
     \frac{R(\varphi)}{R_0}
   \right)
 \right]R(\varphi)^2
 \notag \\
  = & \frac{1}{2}\gamma_0\gamma_1R_0^2e^{-4\kappa\varphi/\sqrt6-1-\frac{2}{\gamma_1}+2W(G(\varphi))}
  \left(
    \frac{1}{2}+W(G(\varphi))
  \right).
\end{align}

Here we regard the scalaron field as an inflaton. It is assumed that the scalaron potential dominates the energy of the early universe and the contribution from the SM particles is neglected. The observed fluctuations of the cosmic microwave background (CMB) constrain the shape of the scalaron potential. We employ the slow roll scenario and fix the parameters, $\gamma_0$, $\gamma_1$ and $R_0$, to satisfy the curvature power spectrum $A_s$, the spectral index $n_s$, tensor-to-scalar ratio $r$, observed in the fluctuations of CMB. After some numerical calculations, a set of consistent parameters is obtained. If we tune the parameters as
\begin{align}
  &\gamma_0 = 1.95 \times10^9, \label{gamma0}\\
  &\gamma_1 = 6.40 \times10^{-3}, \label{gamma1}\\
  &\kappa^2 R_0 = 4.00, \label{R0}
\end{align}
the CMB fluctuations are predicted as
\begin{align}
  &A_s= 2.35 \times10^{-9} ,\\
  &n_s= 0.964 ,\\
  &r= 2.01\times10^{-3}.
\end{align}
These values are consistent with the observed results\cite{Aghanim:2018eyx}. As is pointed out in Ref.~\citen{Odintsov:2017hbk}, the renormalization group flow gives an additional constraint. It should be noted that  we do not adapt to the renormalization group analysis in the present model.


\section{Scalaron as Dark Matter}

\subsection{Constraints from DE}
In the current universe a typical curvature, $R$, is much smaller than the primordial inflation scale, $R_0$. In the DE dominant era, $R_0 \gg R\sim R_C \sim \Lambda_{DE} > 0$, the contribution from the third term in Eq.~(\ref{logF(R)}) should be ignored. We fix $\gamma_0$, $\gamma_1$ and $\kappa^2 R_0$ by Eqs.~(\ref{gamma0}), (\ref{gamma1}) and (\ref{R0}) then the logarithmic correction in the third term is dropped. Thus the model parameters should be assumed to satisfy
\begin{align}
  R_C \gg \kappa^2\gamma_0  R_C^2,
\\
  \Lambda_{DE} \gg \kappa^2\gamma_0 R_C^2,
\\
  |\alpha| \Lambda_{DE} \gg \kappa^2\gamma_0  R_C^2.
\end{align}
Under these assumptions in the DE dominant era, $F(R)$ is approximated by  
\begin{align}
  F (R) & \simeq R - \Lambda_{{DE}} \left[ 1 - \alpha \frac{R}{R_C} \ln
  \left( \frac{R}{R_C} \right) \right] . \label{F:DE}
\end{align}

\begin{figure}[htbp]
 \begin{center}
  \includegraphics[width=100mm]{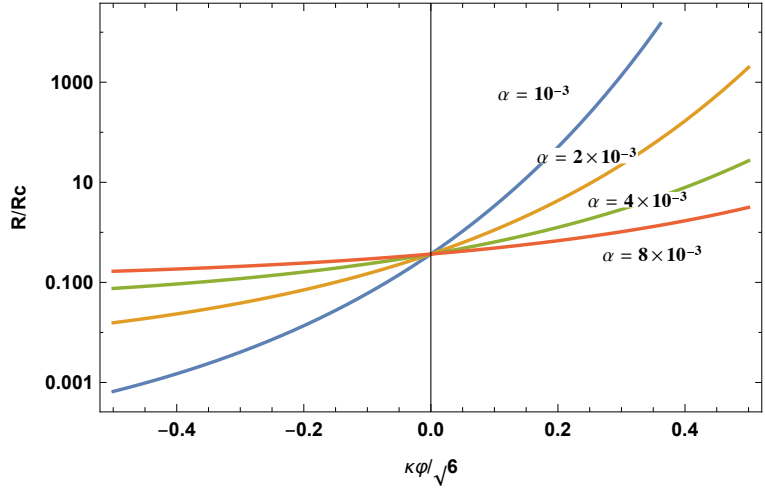}
 \end{center}
 \caption{Curvature $R$ as a function of the scalaron field, $\varphi$, for $R_C/ \Lambda_{DE}=10^{-2}$.}
 \label{R-phi}
\end{figure}

Substituting Eq.~(\ref{F:DE}) into Eq.~(\ref{Rtophi}), we obtain the following relationship between the scalaron field, $\phi$, and the curvature, $R$,
\begin{align}
  R/R_C =\exp \left[ \frac{R_C}{\alpha \Lambda_{{DE}}} (e^{2
  \kappa\varphi/\sqrt6} - 1) - 1 \right] .
\end{align}
Since no stable ground state is observed for a negative $\alpha$ in our numerical calculations, we assume that the $\alpha$ has a positive value below. As is shown in Fig.~\ref{R-phi}, we observe that the curvature, $R$, is monotonically increasing as a function of the scalaron field, $\varphi$. Since we assume $R_0 \gg R\sim R_C \sim \Lambda_{DE} > 0$ in the DE dominant era, a larger $\varphi$ is beyond the approximation (\ref{F:DE}).

\subsection{Scalaron potential}
In the DE dominant era the energy-momentum tensor of SM also contributes the potential energy for the scalaron field. Inserting Eq.~(\ref{F:DE}) into Eq.~(\ref{epot}) with Eq.~(\ref{pot}), we obtain the effective potential,
\begin{align}
  V_{{eff}} (\varphi)
  = \frac{\alpha \Lambda_{{DE}} e^{- 4 \kappa\varphi/\sqrt6}}{2 \kappa^2} \left(
  \exp \left[ \frac{R_C}{\alpha \Lambda_{{DE}}} (e^{2 \kappa\varphi/\sqrt6} - 1) - 1
  \right] + \frac{1}{\alpha} - \frac{\kappa^2 {T^{\mu}}_{\mu}}{2
  \alpha \Lambda_{{DE}}} \right) .
  \label{effpot}
\end{align}
The stable ground state is found by observing the minimum of the effective potential. 

To show the existence of a stable ground state we consider the large and small $\varphi$ limits of the effective potential (\ref{effpot}). For $\kappa\varphi/\sqrt6\gg1$ the effective potential (\ref{effpot}) is simplified to
\begin{align}
  V_{eff}(\varphi)\simeq
  \frac{\alpha\Lambda_{DE}}{2\kappa^2}\exp\left[
  \frac{R_C}{\alpha\Lambda_{DE}}(e^{2\kappa\varphi/\sqrt6}-1)
  \right].\label{lagsigpot}
\end{align}
Since $\alpha$ is assumed to be a positive value, the effective potential  is exponentially increases as a function of $\varphi$. At the small $\varphi$ limit, $\kappa\varphi/\sqrt6\ll1$, the effective potential (\ref{effpot}) reduces to
\begin{align}
  V_{eff}(\varphi)
\simeq
  \frac{\Lambda_{DE}}{2\kappa^2}\left[
 \alpha e^{-1}+1-\frac{\kappa^2{T^\mu}_{\mu}}{2\Lambda_{DE}}+
  \left(
  -2\alpha e^{-1}-2+\frac{e^{-1}R_C}{\Lambda_{DE}}+\frac{\kappa^2{T^\mu}_{\mu}}{\Lambda_{DE}}
  \right)\frac{2\kappa\varphi}{\sqrt6}
  \right].\label{smsigpot}
\end{align}
The effective potential (\ref{smsigpot}) is a decreasing function for
\begin{align}
-2\alpha e^{-1}-2+\frac{e^{-1}R_C}{\Lambda_{DE}}+\frac{\kappa^2{T^\mu}_{\mu}}{\Lambda_{DE}} < 0 .
\label{cond:sg}
\end{align}
In the DE dominant era it is  assumed that $R\sim R_C \sim \Lambda_{DE} > 0$. In the solar system a representative scale of ${T^\mu}_{\mu}$ is $-10^{-17} \mbox{GeV}^4$. Then the major contribution of the left hand side in Eq.~(\ref{cond:sg}) comes from the fourth term which is negative and of the order of ${\kappa^2{T^\mu}_{\mu}}/{\Lambda_{DE}}\sim -10^{30}$.
In this case the effective potential decreases and increases for a small and large $\varphi$, respectively. It shows the existence of a stable ground state.

\begin{figure}[htbp]
	\begin{center}
		\includegraphics[width=100mm]{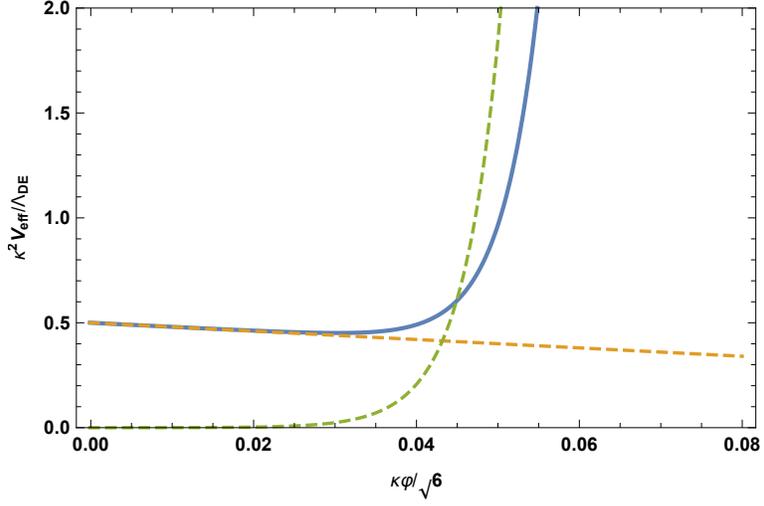}
  \end{center}
	\caption{Behavior of the effective potential (\ref{effpot}) for $\alpha=10^{-4}$, $R_C/ \Lambda_{DE}=10^{-2}$ and $ \kappa^2{T^\mu}_\mu/\Lambda_{DE}=-10^{-5}.$ The green and orange dashed lines represent the approximate equations (\ref{lagsigpot}) and (\ref{smsigpot}), respectively.}
	\label{depapp}
\end{figure}
In Fig.~\ref{depapp} we plot typical behavior of the effective potential (\ref{effpot}). It is observed that the effective potential exponentially increases for $\kappa\varphi/\sqrt{6} \gtrsim 0.04$. The slope is consistent with the approximate equation (\ref{lagsigpot}) (green dashed line). As is shown in Fig.~\ref{depapp} the effective potential (solid blue line) is coincident with the approximate equation (\ref{smsigpot}) (orange dashed line) at the small $\varphi$ limit. A stable ground state is found at $\kappa\varphi/\sqrt{6} \sim 0.03$.

\begin{figure}[htbp]
	\begin{center}
		\includegraphics[width=100mm]{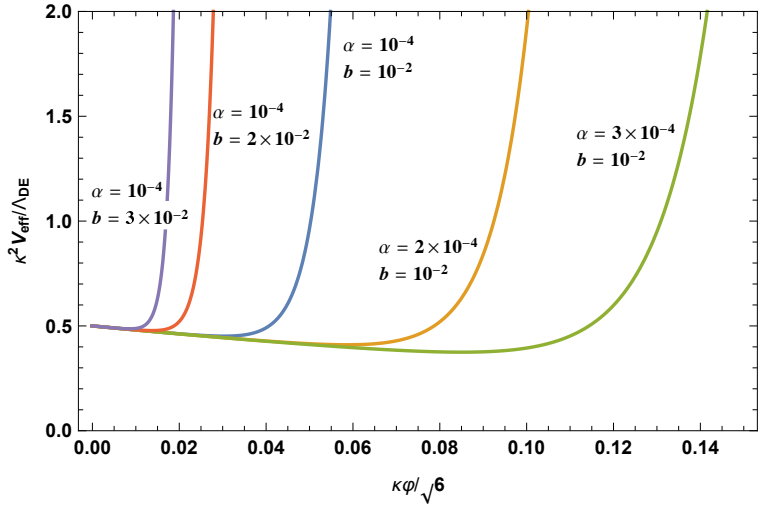}
	\end{center}
	\caption{Behavior of the effective potential for $\kappa^2{T^\mu}_\mu/\Lambda_{DE}=10^{-5}$ as $\alpha$ and $b$ vary.}
	\label{dep}
\end{figure}
To study the contribution from the logarithmic correction term in Eq.~(\ref{F:DE})we plot the effective potential (\ref{effpot}) as $\alpha$ and $b\equiv R_C/\Lambda_{DE}$ vary. It is clearly observed that the logarithmic correction plays a decisive role for the start point of the exponential increasing behavior. Thus the scalaron mass which is given by the second derivative at the minimum significantly depends on the logarithmic correction term. A heavier scalaron mass is generated for smaller $\alpha$ and larger $b$.  
 

The minimum of the effective potential is found by solving the gap equation,
\begin{align}
  \left. \frac{d V_{eff}(\varphi)}{d \varphi}\right|_{\varphi = \varphi_{min}} = 0.
\end{align}
After some calculations, we obtain
\begin{align}
  \kappa\varphi_{min}/\sqrt6=\frac12\ln\left\{\frac{\alpha\Lambda_{DE}}{R_C}\left[2+
  W\left(\frac{1}{\alpha}e^{-1+\frac{Rc}{\alpha\Lambda_{DE}}}\left(2-\frac{\kappa^2{T^\mu}_{\mu}}{\Lambda_{DE}}\right)\right)\right]\right\}.
  \label{phi-min}
\end{align}

\begin{align}
  \kappa\varphi_{min}/\sqrt6\simeq
  \frac12\ln\left\{1+\frac{\alpha\Lambda_{DE}}{R_C}
  \left[
  1+
  \ln\left(\frac{2\Lambda_{DE}-\kappa^2{T^\mu}_{\mu}}{\alpha R_C}\right)
  \right]
  \right\}. \label{miniphi-app}
\end{align}
%
Calculating the second derivative of effective potential (\ref{effpot}) and inserting the field variable (\ref{phi-min}), we obtain the scalaron mass,
\begin{align}
  m_\varphi^2 &= \left. V_{eff}''(\varphi)\right |_{\varphi=\varphi_{min}} \nonumber\\
  &= \frac{(2\Lambda_{DE}-\kappa^2{T^\mu}_\mu) R_C^2}{6\alpha^2\Lambda^2_{DE}}
  \left[
    \frac{1}{W \left( \frac{1}{\alpha}e^{-1+\frac{Rc}{\alpha\Lambda_{DE}}}\left(2-\frac{\kappa^2{T^\mu}_{\mu}}{\Lambda_{DE}} \right) \right)}
    \right. \nonumber\\
  & \hspace{24ex}\left.
    +\frac{1}{2+W \left( \frac{1}{\alpha}e^{-1+\frac{Rc}{\alpha\Lambda_{DE}}}\left(2-\frac{\kappa^2{T^\mu}_{\mu}}{\Lambda_{DE}} \right) \right)}
  \right].
	  \label{mass:DM00}
\end{align}
Here the ratio, $\alpha\Lambda_{DE}/ R_C$, should be much smaller than unity, $\alpha\Lambda_{DE}/ R_C \ll 1$,  from the constraint at the inflation era (\ref{ineq:inf}).
In the solar system the trace of the energy-momentum tensor develops a large and negative value,
\begin{align}
{\kappa^2{T^\mu}_{\mu}}/{\Lambda_{DE}}\sim -10^{30}.
\label{energy-momentum:solar}
\end{align}
Adapting these conditions, we drop non-leading terms and approximate the Lambert $W$-function by,
\begin{align}
  W(x)\simeq \ln x
   \hspace{3mm} \mbox{for} \hspace{3mm} x\gg1.
  \label{W}
\end{align}
and find a simple expression for the scalaron mass,
\begin{align}
  m_\varphi^2 
    \simeq \frac{ (2\Lambda_{DE}-\kappa^2{T^\mu}_\mu) R_C}{3\alpha\Lambda_{DE}} 
  	\frac{1+\frac{\alpha\Lambda_{DE}}{R_C}\ln\left[ \frac{2\Lambda_{DE}-\kappa^2{T^\mu}_\mu}{\alpha\Lambda_{DE}} \right]}
  	{\left( 1+\frac{\alpha\Lambda_{DE}}{R_C} \ln \left[ \frac{2\Lambda_{DE}-\kappa^2{T^\mu}_\mu}{\alpha\Lambda_{DE}} \right] \right)^2 - \left( \frac{\alpha\Lambda_{DE}}{R_C} \right)^2}
	  \label{mass:DM0}
\end{align}
%
\begin{align}
  m_\varphi^2\simeq -\frac{\kappa^2R_C{T^\mu}_{\mu}}{3\alpha\Lambda_{DE}}.
  \label{mass:DM}
\end{align}
The square of the scalaron mass is inversely proportional to the parameter $\alpha$ which is the coefficient of the logarithmic correction term.
Therefore, we conclude that the quantum gravity correction has a significant contribution to the scalaron mass in the DE dominant era. It should be noted that the scalaron mass is not divergent with no logarithmic correction, $\alpha = 0$, because the approximation (\ref{W}) can not be adapted at the limit.

\begin{figure}[H]
 \begin{center}
  \includegraphics[width=100mm]{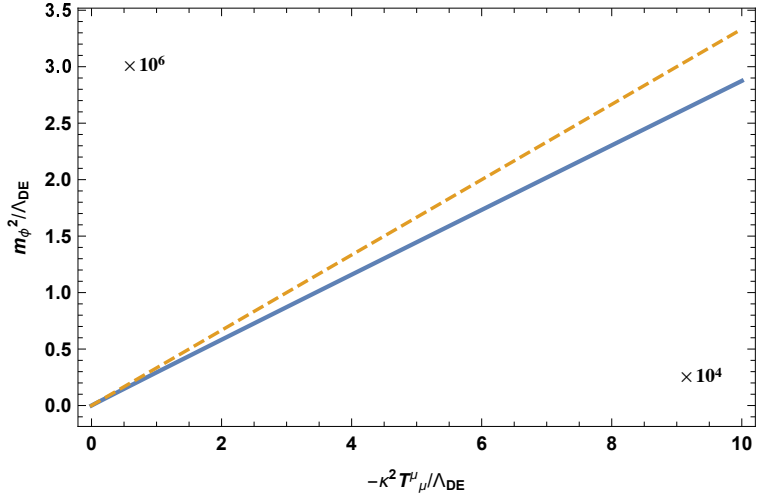}
 \end{center}
 \caption{Scalaron mass as a function of ${T^\mu}_\mu$ for $\alpha=10^{-4}$ and $R_C/ \Lambda_{DE}=10^{-2}$. The solid and dashed lines represent the square of the scalaron mass (\ref{mass:DM00}) and (\ref{mass:DM}), respectively.}
 \label{smass}
\end{figure}

We numerically calculate the square of the scalaron mass (\ref{mass:DM00}) and plot it as a function of ${T^\mu}_\mu$ in Fig.\ref{smass}. The scalaron mass is estimated by Eq.~(\ref{mass:DM}) in the correct order of magnitude.

\subsection{Realistic model parameters}
To identify the scalaron as a dark matter the scalaron lifetime must be longer than the age of the current universe,
\begin{align}
  \tau_{\varphi}>\tau_{\mbox{uni}}\sim10^{17}s.
\end{align}
The decay channel and width of the scalaron depend on the scalaron mass and the interaction with the SM particles. The lower bound of the lifetime imposes the upper bound for the scalaron mass. The scalaron lifetime is roughly estimated by the decay channel into two photons and two gluons through massive fermion and gauge boson loops\cite{Katsuragawa:2016yir, Katsuragawa:2017wge}. Following the discussion in Ref.~\citen{Katsuragawa:2016yir} we obtain the upper bound for the scalaron mass,
\begin{align}
  m_\varphi^2 \lesssim\mbox{O}(1)[\mbox{GeV}^2].\label{upbound}
\end{align}

In the logarithmic F (R) gravity the scalaron mass is given by (\ref{mass:DM}).
Thus the scalaron mass upper bound (\ref{upbound}) guaranteed for 
\begin{align}
  \frac{R_C}{3\alpha\Lambda_{DE}} \lesssim \mbox{O}(10^{55}) ,
 \label{ine:s}
\end{align}
in the solar system, ${T^\mu}_{\mu}\sim -10^{-17} \mbox{GeV}^4$.
This inequality is satisfied for a wide range of $\alpha$ if we set the parameter, $R_C$, at the DE scale, $\Lambda_{DE}$.

\begin{figure}[htbp]
 \begin{center}
  \includegraphics[width=100mm]{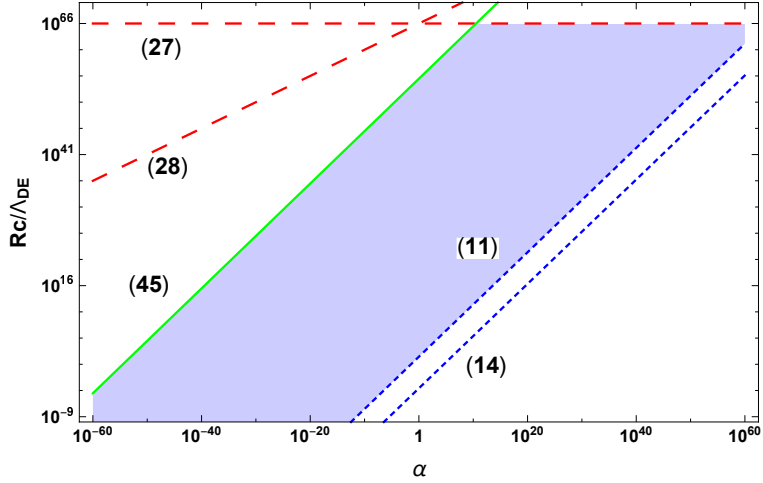}
 \end{center}
 \caption{Allowed region for $\alpha$ and $R_C/ \Lambda_{DE}$.}
 \label{perre}
\end{figure}

Combining the inequality (\ref{ine:s}) with the constraints at inflation era and DE dominant era, we find the allowed value for $R_C/\Lambda_{DE}$ as a function of  $\alpha$. In Fig.~\ref{perre} the colored area indicates the allowed region. Thus the logarithmic F (R) gravity contains a dark matter candidate without fine-tuning the model parameters, $R_C$ and $\alpha$.

\section{Conclusion}
We have studied the scalaron dark matter in the logarithmic $R^2$-corrected $F(R)$ gravity which is constructed by the $R^2$ and cosmological constant terms with the logarithmic corrections. We describe an additional degree of freedom by the scalaron field and identify it as the inflaton at the inflation era and the dark matter at the DE dominant era. The model parameters are tuned to dominate the energy density of the universe by the logarithmic-corrected terms of $R^2$ at the inflation era and cosmological constant at the DE dominant era. 

In the inflation era we adopt the slow roll inflation scenario and constrain the model parameters to reproduce a realistic fluctuation of the cosmic microwave background. Evaluating the effective potential  in the DE dominant era, we find the stable ground state for the scalaron field and calculate the scalaron mass. It is found that the dominant contribution to the scalaron mass comes from the logarithmic correction term and 
the scalaron lifetime is enough long as a dark matter candidate for a wide parameters range in the solar system. Therefore the logarithmic $F(R)$ gravity introduces not only the primordial and current accelerated expansion of the universe but also the gravitational origin of the dark matter.

There are some remaining problems.  It is expected that the model parameters are also constrained in critical phenomena at the early universe \cite{Katsuragawa:2018wbe}. The interaction with the SM particles should be introduced and the reheating process should be analyzed to estimate the relic abundance of the scalaron dark matter. We hope to report some solutions to these problems in the future.


\section*{Acknowledgements}
The authors would like to thank T.~Katsuragawa and S.~D.~Odintsov for
valuable discussions.


\end{document}